\documentclass[journal,twoside]{IEEEtran}

\usepackage[utf8]{inputenc}
\usepackage{amsmath}
\usepackage{dsfont}
\usepackage[tableposition=top]{caption}
\usepackage[labelformat=simple]{subcaption}

\DeclareCaptionLabelSeparator{periodspace}{.\quad}
\usepackage{amssymb}
\usepackage{bm}
\usepackage[colorlinks=true,bookmarks=false,citecolor=blue,urlcolor=blue]{hyperref} 
\usepackage[capitalize]{cleveref}
\crefname{equation}{}{}
\Crefname{figure}{Fig.}{Figs.}
\Crefname{section}{Sec.}{Secs.}
\usepackage[acronym]{glossaries}
\usepackage{adjustbox}
\usepackage{mathtools}
\newtheorem{definition}{Definition}
\newtheorem{operation}{Operation}
\newtheorem{example}{Example}
\usepackage{xspace}

\newcommand{\Rcal}{\mathcal{R}}
\newcommand{\Xcal}{\mathcal{X}}

\newcommand{\Jcal}{\mathcal{J}}

\newcommand{\Rbb}{\mathbb{R}}
\newcommand{\Nbb}{\mathbb{N}}

\newcommand{\Xbb}{\mathbb{X}}

\newcommand{\Hbb}{\mathbb{H}}

\newcommand{\GSS}{\text{GSS}}

\newcommand{\bmx}{\bm{x}}
\newcommand{\bmy}{\bm{y}}
\newcommand{\bml}{\bm{l}}
\newcommand{\bmX}{\bm{X}}
\newcommand{\bmY}{\bm{Y}}
\newcommand{\bmb}{\bm{b}}

\newcommand{\bmC}{\bm{C}}

\newcommand{\argmax}{\text{argmax}}
\newcommand{\RBMD}{\text{R}_\text{BMD}}
\newcommand{\DOF}{\abb{DOF}\xspace}
\newcommand{\DOFs}{\abbpl{DOF}\xspace}

\newcommand{\abb}[1]{\gls*{#1}}

\newcommand{\abbpl}[1]{\glspl*{#1}}

\newacronym{PS}{PS}{probabilistic shaping}
\newacronym{GS}{GS}{geometric shaping}
\newacronym{AIR}{AIR}{achievable information rate}
\newacronym{SSMF}{SSMF}{standard single-mode fiber}
\newacronym{SSFM}{SSFM}{split-step Fourier method}
\newacronym{EDFA}{EDFA}{erbium-doped fiber amplifier}
\newacronym{AWGN}{AWGN}{additive white Gaussian noise}
\newacronym{WDM}{WDM}{wavelength division multiplexing}
\newacronym{PAS}{PAS}{probabilistic amplitude shaping}
\newacronym{MB}{MB}{Maxwell-Boltzmann}
\newacronym[firstplural={degrees of freedom (DOFs)}]{DOF}{DOF}{degree of freedom}
\newacronym{BER}{BER}{bit error rate}
\newacronym{FEC}{FEC}{forward error correction}
\newacronym{GMI}{GMI}{generalized mutual information}
\newacronym{MI}{MI}{mutual information}
\newacronym{NLIN}{NLIN}{nonlinear interference noise}
\newacronym{PM}{PM}{polarization multiplexed}
\newacronym{QAM}{QAM}{quadrature amplitude modulation}
\newacronym{PAM}{PAM}{pulse amplitude modulation}
\newacronym{ASK}{ASK}{amplitude-shift keying}
\newacronym{PCS}{PCS}{probabilistic constellation shaping}
\newacronym{SE}{SE}{spectral efficiency}
\newacronym{GSS}{GSS}{geometric shell shaping}
\newacronym{PAPR}{PAPR}{peak-to-average power rating}
\newacronym{BRGC}{BRGC}{binary reflected Gray code}
\newacronym{BICM}{BICM}{bit-interleaved coded modulation}
\newacronym{DSP}{DSP}{digital signal processing}
\newacronym{4D}{4D}{four-dimensional}
\newacronym{OSNR}{OSNR}{optical signal-to-noise ratio}
\newacronym{SNR}{SNR}{signal-to-noise ratio}
\newacronym{LLR}{LLR}{log-likelihood ratio}
\newacronym{BMD}{BMD}{bit-metric decoding}
\newacronym{SMD}{SMD}{symbol-metric decoding}
\newacronym{HD}{HD}{hard-decision}
\newacronym{SD}{SD}{soft-decision}
\newacronym{SCC}{SCC}{staircase code}
\newacronym{BCH}{BCH}{Bose-Chaudhuri-Hocquenghem}
\newacronym{NL}{NL}{nonlinear}
\newacronym{PDF}{PDF}{probability density function}
\newacronym{PMF}{PMF}{probability mass function}
\newacronym{RRC}{RRC}{root-raised cosine}
\newacronym{CD}{CD}{chromatic dispersion}
\newacronym{MDM}{MDM}{multidimensional modulation}

\begin{document}

\title{Introducing 4D Geometric Shell Shaping for Mitigating Nonlinear Interference Noise}

\author{
    Sebastiaan Goossens, \textit{Student Member, IEEE},
    Yunus Can Gültekin, \textit{Member, IEEE},\\
    Olga Vassilieva,  \textit{Senior Member, IEEE},
    Inwoong Kim, \textit{Senior Member, IEEE},\\
    Paparao Palacharla, \textit{Senior Member, IEEE},
    Chigo Okonkwo, \textit{Senior Member, IEEE},\\
    and Alex Alvarado, \textit{Senior Member, IEEE}
    \thanks{S. Goossens, Y. C. Gültekin and A. Alvarado are with the Information and Communication Theory Lab, Signal Processing Systems Group, Department of Electrical Engineering, Eindhoven University of Technology, Eindhoven 5600 MB, The Netherlands (e-mails: \{s.a.r.goossens, y.c.g.gultekin, a.alvarado\}@tue.nl).}
    \thanks{C. M. Okonkwo is with the High Capacity Optical Transmission laboratory, Eindhoven Hendrik Casimir Institute (EHCI), Department of Electrical Engineering, Eindhoven University of Technology, Eindhoven 5600 MB, The Netherlands (e-mail: c.m.okonkwo@tue.nl).}
    \thanks{O. Vassilieva, I. Kim and P. Palacharla are with the Fujitsu Network Communications, Inc., Richardson, 75082 TX, USA (e-mails: \{olga.vassilieva, inwoong.kim, paparao.palacharla\}@fujitsu.com).}
    \thanks{The work of S. Goossens, Y. C. Gültekin and A. Alvarado has received funding from the European Research Council (ERC) under the European Union’s Horizon 2020 research and innovation programme (grant agreement No 757791) and via the Proof of Concept grant (grant agreement No 963945).}
    \thanks{Parts of this paper were presented at Signal Processing in Photonic Communications (SPPCom), Maastricht, The Netherlands, July 2022.}
}

\markboth{Preprint}{Goossens \MakeLowercase{\textit{et al.}}: Introducing 4D Geometric Shell Shaping for Mitigating Nonlinear Interference Noise
}

\maketitle

\begin{abstract} 
Four dimensional geometric shell shaping (4D-GSS) is introduced as an approach for closing the nonlinearity-caused shaping gap. This format is designed at the spectral efficiency of 8 bit/4D-sym and is compared against polarization-multiplexed 16QAM (PM-16QAM) and probabilistically shaped PM-16QAM (PS-PM-16QAM) in a 400ZR-compatible transmission setup with high amount of nonlinearities. Reach increase and nonlinearity tolerance are evaluated in terms of achievable information rates and post-FEC bit-error rate. Numerical simulations for a single-span, single-channel show that 4D-GSS achieves increased nonlinear tolerance and reach increase against PM-16QAM and PS-PM-16QAM when optimized for bit-metric decoding ($\RBMD$). In terms of $\RBMD$, gains are small with a reach increase of 1.7\% compared to PM-16QAM. When optimizing for mutual information, a larger reach increase of 3\% is achieved compared to PM-16QAM. Moreover, the introduced GSS scheme provides a scalable framework for designing well-structured 4D modulation formats with low complexity.
\end{abstract}


\section{Introduction}

\IEEEPARstart{I}{n} recent years, constellation shaping techniques such as \abb{PS} \cite{Bocherer2015, Buchali2016, Fehenberger2016, Amari2019, Goossens2019, Fehenberger2020} and \abb{GS} \cite{Qu2017,Zhang2017,Chen2018,Chen2019,Chen2021} have been widely investigated to cope with the exponentially increasing capacity demand of optical fiber communications. These techniques can be used to alter the properties of the transmitted constellation to increase the \abbpl{AIR} of a communication system.
\abb{PS} imposes a nonuniform probability distribution on the constellation points of a square constellation while \abb{GS} changes the location of the constellation points and allows for nonequidistant spacing of the points. Combinations of PS and GS are also increasingly investigated in recent literature \cite{Cai2018,Ding2021}. This combination is called hybrid shaping.  It has been shown that for the \abb{AWGN} channel and a given (finite) constellation cardinality, \abb{PS} outperforms \abb{GS} \cite[Fig. 2]{Chen2018}, \cite[Figs. 2, 3]{Steiner2017} and also allows for very fine rate-adaptivity \cite[Fig. 1]{Bocherer2015}. This comes however at the cost of requiring additional steps in the \abb{DSP} chain. For example, if the \abb{PAS} architecture \cite{Bocherer2015} is used, a shaper and a deshaper are required. \abb{GS} has the advantage of not requiring extra \abb{DSP}, but it does require changes to the mapper and demapper. It is well known that demapper complexity increases when increasing demapping dimensionality \cite{Fuentes2016}, however, it has been shown that careful design can achieve a good balance between performance and complexity for multidimensional soft demappers (see e.g., \cite{Yoshida2016}). In \cite{Fuentes2016}, one method of reducing the complexity in the demapper is by exploiting quadrant symmetry in 2D constellations. This results in a reduction of the number of required Euclidean distance calculations by around a factor of four with minimal performance loss. This reduction factor is expected to scale exponentially with the number of dimensions.

Designing constellations via \abb{GS} lifts restrictions previously in place from equally-spaced square \abb{QAM}. Conventional shaped constellation design is done by targeting the \abb{AWGN} channel, for which a Gaussian-like constellation shape is optimal \cite[Chs. 8, 9]{Cover2005}. The resulting increase in performance is called linear shaping gain. \abb{PS} on square (2D) \abb{QAM} achieves this by targeting the same \abb{MB} distribution on every dimension, while \abb{GS} places constellation points in the 2D I-Q plane  allowing nonuniform distance between the constellation points. In the context of optical communications, \abb{AWGN}-optimized constellations have been applied in nonlinear optical fiber communications \cite{Buchali2016, Chen2018, Qu2019, Cho2019} by using polarization multiplexing.
While this method of designing constellations is relatively simple and provides considerable performance improvements \cite{Fehenberger2016}, the nonlinear nature of optical fibers results in an increasing mismatch between the assumed channel model and targeted optical link for increasing launch powers \cite{Renner2017}. 

To further improve performance in fiber-optical systems, research has moved towards designing constellations with nonlinear tolerance in mind \cite{Renner2017,Chen2021_2}. 
It has been shown that multidimensional constellation design is able to provide tolerance against nonlinear fiber effects by reducing \abb{NLIN}, effectively closing the nonlinearity-caused shaping gap, in addition to providing linear shaping gains \cite{Kojima2017,Chen2019}. One explanation is that the set of multidimensional constellations that can be expressed as the Cartesian product of a lower-dimensional constellation with itself is a subset of all possible multidimensional constellations. Thus, optimizing the multidimensional constellation, instead of a lower-dimensional constellation, results in larger potential gains \cite[Fig. 3(c)]{Dar2014}, \cite[Fig. 1]{Chen2022}, \cite{Goossens2022}.

Multiple forms of multidimensional modulation exist in the literature. Extending the constellation design to utilize both polarizations jointly in a single time instance is well known (e.g., \cite{Chen2019}). This is called 4D modulation throughout this paper. Instead of employing polarization multiplexing, 4D modulation focuses on the design of constellations that are not a Cartesian product of lower-dimensionality constellations. Other methods include extending the number of dimensions over multiple time-slots \cite{Borowiec2014,Chen2019_2} or multiple (sub)-carriers \cite{Kojima2018}. While the above-mentioned methods are based on \abb{GS}, short-blocklength \abb{PS}, which can be considered a form of multidimensional modulation over multiple time-slots, also provides tolerance to nonlinearities \cite{Amari2019, Skvortcov2021}.

The challenge in designing constellations with more than two dimensions is the exponential increase in \DOFs that is generally associated with increasing the number of dimensions while maintaining equal \abb{SE} per real dimension.
While simple unconstrained optimizations provide the best theoretical possible performance, they are time consuming, might not lead to a global optimum, and generally provide unstructured results.
From a practical point of view, well-structured constellations are always preferred since they allow for more efficient demapping strategies (e.g., by utilizing separability and symmetry of the constellations) \cite{Nakamura2018,Wang2020}.

Designing an optimal constellation generally requires some form of iterative optimization for evaluating performance after each iteration. A channel model of the target communication link is thus required. The most straightforward method is using the computationally expensive but very accurate \abb{SSFM} method. It is however desirable to reduce complexity of the optimization procedure to a more manageable level by using a simplified model or a simplified optimization objective. For simplified models, closed-form models which approximate the \abb{NLIN} are available, like the EGN model \cite{Carena2014}, or its recently introduced extension to dual-polarization \cite{Liga2022}. Model-aided optimization is used, for example, in \cite{Renner2017} and \cite{Sillekens2018}, where the EGN model, or derivations thereof, are used to simplify the evaluation of the optical channel during the optimization stage.

Currently, 4D geometrically shaped constellations optimized under an \abb{AWGN} channel assumption providing reach increase for optical communications exist up to 10 bits/4D-sym \cite[Fig. 2]{Chen2021_2}. Similar constellations designed under an optical channel assumption also exist up to 10 bits/4D-sym \cite[Fig. 3]{Chen2021_2} and 12 bits/4D-sym \cite{Oliari2022}, which employ machine learning to cope with the optimization complexity.

Another way of simplifying the design is reducing the \DOFs within the optimization problem. This reduction in \DOFs is generally achieved by imposing constraints which exploit existing regularities (e.g., symmetries), as used in \cite{Qu2018,Chen2021}. However, these works target the \abb{AWGN} channel for constellation design, which potentially impacts performance negatively when applied to a nonlinear fiber communication system.

In this paper, a novel framework is introduced for geometrically optimizing 4D constellations for the nonlinear fiber channel by using shell constraints together with symmetry constraints. This framework is applied to both \abb{SMD} and \abb{BMD} systems \cite{Fehenberger2016}.
This approach, denoted as 4D \abb{GSS}, provides well-structured constellations, reduces optimization complexity, and leads to negligible performance degradation, all while providing increased nonlinear tolerance associated with multidimensional \abb{GS} optimization.
4D-\abb{GSS} is studied for a \abb{SE} of 8 bits/4D-sym such that the proposed format can find application to the 400ZR system \cite{OIF-400ZR} (or the 800ZR system under preparation), which uses PM-16QAM for transmission. 
To the best of our knowledge, 4D geometrically-shaped constellations with the same \abb{SE} as PM-16QAM specifically designed for nonlinear fiber transmission and using \abb{BMD} do not exist in the literature. 
This is most likely because PM-16QAM with the \abb{BRGC} is very competitive already in terms of \abbpl{AIR} with \abb{BMD}. 
Nevertheless, our optimized format demonstrate increased nonlinear tolerance, and in this challenging scenario, small reach increases against PM-16QAM and probabilistically shaped PM-16QAM are also reported.

The paper is organized as follows. In \cref{sec:4D-GSS} the design of the \abb{GSS} modulation format is explained. \cref{sec:sys_setup_optim} explains the system setup and optimization. Results are discussed in \cref{sec:results} and conclusions are drawn in \cref{sec:conclusions}.
\section{4D Geometric Shell Shaping}\label{sec:4D-GSS}

 \begin{figure*}[ht!]
	\centering
	\begin{adjustbox}{width=0.7\textwidth}
         \includegraphics[]{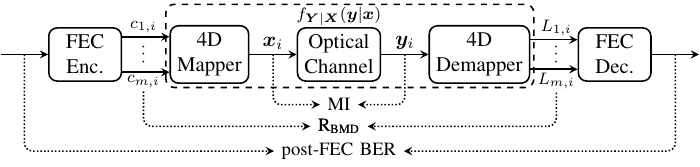}
	\end{adjustbox}
	\caption{System model under consideration. $c_{k,i}$ and $L_{k,i}$ are the coded bits and LLRs for the $i^\text{th}$ symbol, $f_{\bmX|\bmY}(\bmy|\bmx)$ denotes the channel law.}
	\label{fig:system_setup_standard}
\end{figure*}

\subsection{Notation Convention} 
Calligraphic letters $\Xcal$ represent sets. Blackboard bold letters $\Xbb$ denote matrices in which $\bmx_i$ are row vectors denoting the $i^\text{th}$ row. Conditional probability density functions (PDFs) are denoted by $f_{\bmY|\bmX}(\bmy|\bmx)$, where $\bmY$ and $\bmX$ denote random (4D) vectors and $\bmy$ and $\bmx$ denote their realizations. Probability mass functions (PMFs) are denoted by $P_{\bmX}(\bmx)$, expectations are denoted by $E[\cdot]$ and $\overline{(\cdot)}$ denotes binary negation. The squared Euclidean norm of a matrix is denoted by $||\Xbb||^2 = ||\bmx_1||^2+\ldots+||\bmx_M||^2$, where $||\bmx_i||^2 = x_{1i}^2+\ldots+x_{4i}^2$. The indicator function is denoted by $\mathds{1}[\cdot]$, which is 1 when its argument is true and 0 otherwise, and $\Rbb$ and $\Nbb$ denote the set of all real and natural numbers, respectively, with $\Rbb_{>0}$ denoting the set of all positive real numbers.

Throughout this paper, we consider a constellation with $N=4$ real dimensions denoted by the $4 \times M$ real-valued matrix $\Xbb = [\bmx_1,\bmx_2,\ldots,\bmx_M]^\top$, where $M=2^m$ is the constellation cardinality, $\bmx_i\triangleq (x_{1i},x_{2i},x_{3i},x_{4i})\in\Rbb^4$ is a vector denoting the $i$-th constellation point of $\Xbb$ with $\bmx_i \neq \bmx_j$ for $i\neq j$. The 2D coordinates of the x- and y-polarization are represented by $x_{1i},x_{2i}$ and $x_{3i},x_{4i}$, respectively, and the rows of $\Xbb$ are labeled using a fixed binary labeling. In other words, the $i$-th symbol $\bmx_i$ is associated with a unique binary label $\bmb_i=(b_{1i},b_{2i},\ldots,b_{mi})$.

\subsection{Achievable Information Rates}
A number of metrics exist for evaluating the performance of a fiber optical system. For systems based on \abb{SMD}, the \abb{MI} is often used. Using Monte-Carlo simulations, the \abb{MI} can be approximated as
\begin{equation}\label{eq:MI}
    \begin{aligned}
        \text{MI} \triangleq I(\bmX;\bmY) \approx{}& \frac{1}{D}\sum_{i=1}^D \log_2 \frac{q_{\bmY|\bmX}(\bmy_i|\bmx_i)}{\sum_{j=1}^M P_{\bmX}(\bmx_j) q_{\bmY|\bmX}(\bmy_i|\bmx_j)},
    \end{aligned}
\end{equation}
where $P_{\bmX}(\bmx_j)$ is the probability of the symbol $\bmx_j$, $D$ is the number of transmitted symbols and $q_{\bmY|\bmX}$ is the auxiliary channel, which is an approximation of the actual channel law $f_{\bmY|\bmX}$ from which the samples $\bmy_1,\bmy_2,\ldots,\bmy_D$ are taken from. We use mismatched decoding \cite{Lapidoth1994} in this paper for which $q_{\bmY|\bmX}$ in \eqref{eq:MI} is considered to be the \abb{AWGN} channel, i.e.,
\begin{equation}\label{eq:PDF_Gauss}
    q_{\bmY|\bmX}(\bmy|\bmx) = \frac{1}{(\pi\sigma^2/2)^2}\exp{\left(-\frac{||\bmy-\bmx||^2}{\sigma^2/2}\right)},
\end{equation}
where $\sigma^2$ is the total noise variance of the 4D \abb{AWGN} channel.
The expression in \eqref{eq:MI} is a modified version of \cite[Eq. (30)]{Alvarado2018} that takes probabilities into account (for \abb{PS}). 

For \abb{BICM} systems with \abb{BMD} and probabilistic shaping, an \abb{AIR} is the $\RBMD$ \cite[eq. (1)]{Bocherer2014},\cite[eq. (6)]{Fehenberger2016}. When bit levels are independent, which is the case for conventional uniform signaling, the BMD rate reduces to the well known generalized MI \cite[Thm. 4.11, Coroll. 4.12]{Szczecinski2015},\cite{Fabregas2010}.

We approximate the BMD rate via Monte-Carlo simulations as \cite[Eq. (8)]{Fehenberger2016}
\begin{align}\label{eq:RBMD_def}
    \RBMD &\triangleq \sum_{k=1}^m I(C_k;\bmY) - \sum_{k=1}^m H(C_k) + H(\bmC) \\
    &~\begin{aligned}\label{eq:RBMD_mc}
        \approx{}&-\sum_{j=1}^MP_{\bmX}(\bmx_j)\log_2 P_{\bmX}(\bmx_j) \\ &-\frac{1}{D}\sum_{k=1}^m\sum_{i=1}^D\log_2\left(1+e^{(-1)^{c_{k,i}}L_{k,i}}\right),
    \end{aligned}
\end{align}
where $C_k$ is the random variable representing the transmitted bit at bit position $k$, $I(C_k;\bmY)$ is the bit-wise \abb{MI} between $C_k$ and output $\bmY$, $c_{k,i}$ are the transmitted coded bits and $L_{k,i}$ are the \abbpl{LLR} defined as 
\begin{equation}\label{eq:LLR}
    L_{k,i}\triangleq\log\frac{\sum_{ j\in\Jcal^k_1}q_{\bmY|\bmX}(\bmy_i|\bmx_j)P_{\bmX}(\bmx_j)}{\sum_{ j\in\Jcal^k_0}q_{\bmY|\bmX}(\bmy_i|\bmx_j)P_{\bmX}(\bmx_j)},
\end{equation}
where $\Jcal_b^k$ is the set of constellation point indices with $b\in\{0,1\}$ at bit position $k$. 

To evaluate the performance predictions made by the above-mentioned \abbpl{AIR}, post-\abb{FEC} \abbpl{BER} will also be calculated. The system model is shown in \cref{fig:system_setup_standard} indicating the performance metrics which are considered in this paper. 
 
\subsection{Optimizing Geometric Shaping in 4D}\label{subseq:optim_GSS_4D}
The constellation $\Xbb$ is typically designed to maximize a certain performance metric \cite{Chen2018}. In this paper, $\RBMD$ is the chosen metric and the optimal constellation is denoted by $\Xbb^*$. The resulting optimization problem is defined as
\begin{equation}
    \Xbb^* = \underset{\Xbb\in\Xcal}{\argmax}~\RBMD(\Xbb),
    \label{eq:argmax_unconstrained}
\end{equation}
where $\RBMD$ is given by \cref{eq:RBMD_mc} and \cref{eq:LLR}, and $\Xcal$ is the set containing all $4\times M$ real-valued matrices satisfying a variance (power) constraint, i.e.,
\begin{equation}\label{eq:Xset}
    \begin{aligned}
        \Xcal \triangleq \{& \Xbb : \bmx_i\in\Rbb^4,i=1,2,\ldots,M,E[||\Xbb||^2]\leq P\}.
    \end{aligned}
\end{equation}

The optimization problem in \cref{eq:argmax_unconstrained} has four \DOFs per constellation point, one for each dimension, resulting in $4M$ \DOFs. We call this the {\it unconstrained optimization} and denote it by 4D-\abb{GS}. The \DOFs are directly related to the dimensionality $4$ and the constellation cardinality $M=2^m$. From this we see that for increasing constellation sizes, the \DOFs grow exponentially and the optimization becomes challenging.

In this paper, instead of solving the unconstrained optimization in \eqref{eq:argmax_unconstrained}, we define a set of constraints which reduce the \DOFs, while minimizing the potential loss in performance. The optimization under these constraints is defined as
\begin{equation}\label{eq:argmax_GSS}
    \Xbb^* = \underset{\Xbb\in\Xcal_\GSS}{\argmax}~\RBMD(\Xbb),
\end{equation}
where the optimization space is constrained to $\Xcal_\GSS \subset \Xcal$. As we will show below, the imposed constraints make the optimization problem in \eqref{eq:argmax_GSS} to only have $28$ \DOFs instead of $1024$ \DOFs for the chosen system $(m=8)$. 

In this paper we propose to impose three constraints on the constellation, which we call (i) ``uniform $t$-shell division'', (ii) ``X-Y symmetry'', and (iii) ``orthant\footnote{An orthant is a generalization in $N$-dimensional Euclidean space of what a quadrant is in the 2D plane.} symmetry''. We call the optimization under these constraints 4D-\abb{GSS}. In what follows, we explain these three constraints and how they lead to $28$ \DOFs for $m=8$. As we will show in \cref{sec:results_RBMD}, the loss in performance by introducing these three constraints is minimal.

\subsection{GSS Constraints}
\begin{figure}
	\centering
	\begin{adjustbox}{width=0.5\textwidth} 
        \includegraphics[width=\linewidth]{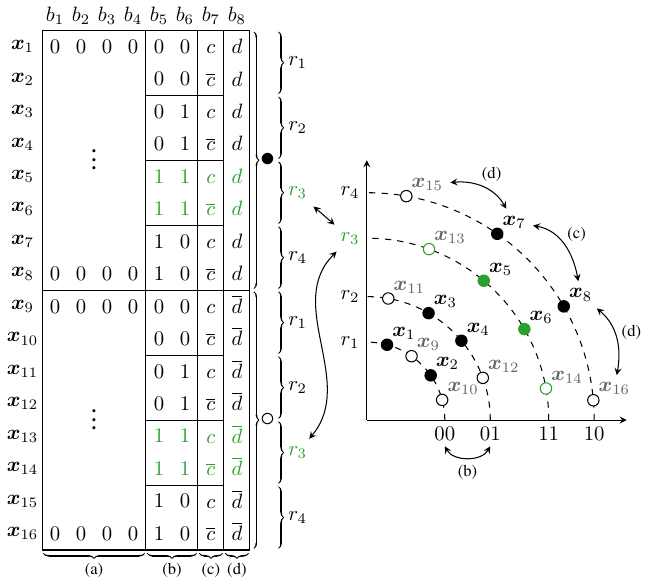}
	\end{adjustbox}
	\caption{Example of binary label bit allocation in the first orthant (left), and an arbitrary 2D projection of the first orthant of a 4D constellation (right), are shown for $m=8$ bits and $t=4$ shells. Shell constraints (bits $b_5$ and $b_6$) and X-Y symmetry (bit $b_8$) are indicated with (b) and (d), respectively. Highlighted in green are the four points on the third shell ($b_5=b_6=1$), and \scalebox{1.25}{$\bullet$} and \scalebox{1.25}{$\circ$} indicate X-Y symmetric points.}
	\label{fig:example}
\end{figure}
Each of the three constraints mentioned in \cref{subseq:optim_GSS_4D} is associated with a part of the binary labeling, which is considered to be fixed. \Cref{fig:example} provides an example of how the bit allocation is predefined under the considered constraints in a 4D constellation with $m=8$ bits and $t=4$ shells. The left side of \cref{fig:example} shows 16 constellation points belonging to a single orthant with their corresponding binary labels. The binary labels are grouped into sets of bits referred to by (a) through (d). The four bits in (a) define an orthant. In \cref{fig:example} only the first orthant is shown and thus, all $\bmx_i$ have the same binary label (all zeros) for (a). The two bits in (b) determine the shell, with each shell having the same amount of points ($2$ in this case). The bits in (c) select between two points on a shell, and (d) selects between two X-Y symmetric points.

The three GSS constraints described above reduce the number of 4D constellation points to be optimized from $2^8=256$ to only $2^{m-5}=8$ (filled circles in Fig.~\ref{fig:example}) with a reduction in \DOFs from $1024$ to $28$. In what follows, we formally describe the set $\Xcal_\GSS$ together with two symmetry operations that make up these three GSS constraints. 

\begin{definition}[Uniformly divided t-shell constraint]\label{def:shell_constr}
    \begin{subequations}
        \begin{align}
            \Xcal_\GSS \triangleq \{&\Xbb : \bmx_i\in\Rbb_{>0}^4, ||\bmx_i||\in\Rcal_t,\\
            & \sum_i \mathds{1}\left[||\bm{x}_i||=r_j\right] = \frac{2^{m-5}}{t},\label{eq:X_GSS_b}\\
            &i=1,2,\ldots,2^{m-5}, j=1,2,\ldots,t \},
        \end{align}
    \end{subequations}
    where
    \begin{equation}\label{eq:R}
        \Rcal_t \triangleq \{ r_1,r_2,\ldots,r_t : t = 2^p < 2^{m-5}, p\in\Nbb\},
    \end{equation}
     and $r_j$ is the radius of the $j^\text{th}$ 4D shell out of a total of $t$ 4D shells.
\end{definition}

The shell constraint forces $2^{m-5}$ constellation points to be equally divided on the $t$ concentric 4D shells in the all positive $\Rbb^4_{>0}$ space (first orthant). This \textit{uniformly divided} part of this constraint is provided by \cref{eq:X_GSS_b}.
The upper limit for $t$ is given by $2^{m-5}$ (see \eqref{eq:R}), which is equivalent to having a dedicated shell for each constellation point. In the case of $t=1$, this constraint turns into a constant modulus constraint, effectively creating a generalization of the format proposed in \cite{Chen2019}.
By forcing each point to be on top of a certain shell, we take away one additional \DOF per constellation point, such that $3$ \DOFs per constellation point remain. However, $t$ extra \DOFs are added due to the number of the shells. This results in $3\cdot2^{m-5}+t$ \DOFs.
The advantage of having an integer power of two ($t=2^p$) shells is that $p$ out of $m-5$ bits can be used to select the shell. In \cref{fig:example}, $p=2$. This offers the possibility of achieving rate-adaptivity by adding \abb{PS} on top of \abb{GSS} using the \abb{PAS} architecture \cite{Bocherer2015}, which is called hybrid shaping. In this case, only the bits which select the shell are shaped. The remaining $m-p-5$ uniform bits select the specific constellation points on a shell.

In the remainder of this section we will assume an identical setup to the one in the example in \cref{fig:example}. As a result $|\Xcal_\GSS|=8$ and the $8$ constellation points are labeled by $\bml_i = [b_{5i},b_{6i},b_{7i}]\in\{0,1\}^3$.

\begin{operation}[X-Y symmetry]\label{op:XY_symm}
    An X-Y symmetry operation applied to 8 points $\bmx_i$ with $i = 1,2,\ldots,8$ results in 16 points and binary labels
    \begin{equation}
        \arraycolsep=1pt
        \begin{array}{lrlr}
            \bmx_{i} &= [x_{1i},x_{2i},x_{3i},x_{4i}], &\quad\widetilde{\bml}_i&=[\bml_i,b_8], \\
            \bmx_{i+8} &= [x_{3i},x_{4i},x_{1i},x_{2i}], &\quad\widetilde{\bml}_{i+8}&=[\bml_i,\overline{b_8}].
        \end{array}
    \end{equation}
\end{operation}

X-Y symmetry mirrors the points in the $\Xcal_\GSS$ set over its two polarizations. A single bit added to the labeling end is used to distinguish between the two X-Y symmetric points.
The X-Y symmetry also ensures identical average transmit power over the two polarizations. In Fig.~\ref{fig:example}, this mirroring causes for example $\bmx_7$ to become $\bmx_{15}$ and the extra bit added to the binary labels is $d$ and $\overline{d}$, resp. After applying the \mbox{X-Y} symmetry operation to $\Xcal_\GSS$, we must apply the orthant symmetry operation, defined as follows.

\begin{operation}[Orthant symmetry]\label{op:orthant_symm}
    The orthant symmetry operation applied to 16 points $\bmx_i$ with $i = 1,2,\ldots,16$ gives
    \begin{equation}\label{eq:OS_definition}
        \arraycolsep=1pt
        \begin{array}{rl}
            \bmx_{i+16(j-1)} &= \bmx_i\Hbb_j, \\
            \bmb_{i+16(j-1)} &= [l_1,l_2,l_3,l_4,\widetilde{\bml}_i],
        \end{array}
    \end{equation}
    for $j = 1,2,\ldots,16$, and where $\Hbb_j$ is the mirroring matrix of the $j$-th orthant
    \begin{equation}\label{eq:OS_mirror_matrices}
        \Hbb_j = \left[
        \arraycolsep=0pt
        \begin{array}{cccc}
            (-1)^{l_1} & 0 & 0 & 0 \\
            0 & (-1)^{l_2} & 0 & 0 \\
            0 & 0 & (-1)^{l_3} & 0 \\
            0 & 0 & 0 & (-1)^{l_4}
        \end{array}
        \right],
    \end{equation}
    where $[l_1,l_2,l_3,l_4]$ is the binary representation of $j$, i.e., ${j-1=\sum_{k=1}^4 l_k 2^{k-1}}$ with $l_k \in \{0,1\}$.
    
    The mirroring matrices transform each of the 16 points in $\bmx_i$ to all $2^N=16$ orthants with corresponding binary labels $\bmb_i$, resulting in a total of 256 constellation points and label combinations, given by
    \begin{equation}\label{eq:OS_example}
        \arraycolsep=0pt
        \begin{array}{lrrrrrlr}
            \bmx_{i}&=[&x_{1i},&x_{2i},&x_{3i},&x_{4i}],&~\bmb_{i}&=[0,0,0,0,\widetilde{\bml}_i], \\
            \bmx_{i+16}&=[&-x_{1i},&x_{2i},&x_{3i},&x_{4i}],&~\bmb_{i+16}&=[1,0,0,0,\widetilde{\bml}_i], \\
            \bmx_{i+32}&=[&x_{1i},&-x_{2i},&x_{3i},&x_{4i}],&~\bmb_{i+32}&=[0,1,0,0,\widetilde{\bml}_i], \\
            \multicolumn{6}{c}{$\vdots$} & \multicolumn{2}{c}{$\vdots$} \\
            \bmx_{i+240}\hspace*{2pt}&=[&-x_{1i},&-x_{2i},&-x_{3i},&-x_{4i}],&~\bmb_{i+240}\hspace*{2pt}&=[1,1,1,1,\widetilde{\bml}_i].
        \end{array}
    \end{equation}
\end{operation}

The first orthant $\Rbb_{>0}^4$ in \cref{def:shell_constr} only considers 4D symbols with all their components to be positive. Orthant symmetry is achieved by mirroring the $2^{m-4}$ constellation points (after applying the X-Y symmetry operation) with respect to the origin along the axes of the $4$ dimensions and by changing the bits $b_1,b_2,b_3,b_4$ (see \cref{eq:OS_definition}), where each of these bits is associated with the sign of one real dimension. An advantage of assigning bits in this manner is that it ensures the orthants themselves are Gray-labeled (adjacent orthants differ only in one bit), which provides higher \abbpl{AIR} compared to other labeling strategies when used in \abb{BICM} systems \cite{Caire1998}. In the context of optical communications, the mirroring procedure was first used in \cite[Sec. II-B]{Chen2021}.

\begin{figure*}[t!]
    \begin{minipage}[t]{0.65\textwidth}
        \includegraphics[]{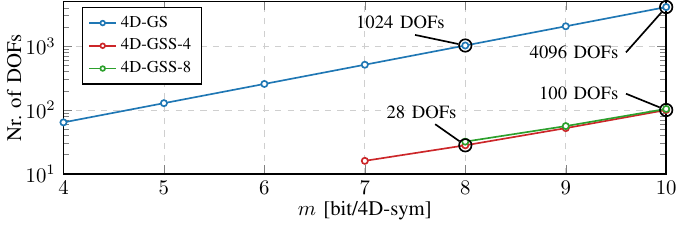}
        \captionof{figure}{Constellation cardinality vs. number of DOFs. The number of DOFs can be used as an indicator of the time it takes for the optimization to converge.}
        \label{fig:DOFs}
    \end{minipage}%
    \begin{minipage}[t]{0.35\textwidth}
            \renewcommand{\arraystretch}{1.1}
            \vspace{-24ex}
            \centering%
            \captionof{table}{DOFs comparison for different\\constellations with $m=8$.}
            \label{tab:params}
            \small
            \begin{tabular}{lccc}
            \hline
            \textbf{Name}  & \textbf{DOFs} & \textbf{4D shells} \\ \hline
            PM-16QAM        & -                   & 9                          \\
            PM-2D-16-GS  & 64                  & 136                  \\
            4D-256-GS  & 1024                  & 256                  \\
            4D-256-GSS-4  & 28                  & 4                  \\
            4D-256-GSS-8  & 32                  & 8                  \\ \hline
            \end{tabular}%
    \end{minipage}
\end{figure*}

\begin{example}[Application to 400ZR]\label{def:shell_constr}
The described constraints are applied to 4D constellations targeting the same \abb{SE} as uniform PM-16QAM, therefore $m=8$. The number of shells is chosen to be $t=4$.
First, the uniformly divided $t$-shell constraint is applied to the $2^m-5=8$ constellation points within the first orthant. This results in the set $\Xcal_\GSS$ with $3\cdot2^{m-5}+t = 28$ \DOFs. Since $t=4$, two bits are used to index the shell. Applying the X-Y symmetry operation to $\Xcal_\GSS$ increases the number of constellation points in the first orthant by a factor 2 to $2^m-4=16$. When applying the orthant symmetry operation, four bits are assigned to select the orthant with groups of two bits assigned to the quadrants in the X and Y polarization respectively, which together define the orthant. This increases the number of constellation points by a factor $2^4$, which results in the desired total amount of $2^8=256$ 4D constellation points.
\end{example}

By using the three constraints described above, the \DOFs are reduced from an unconstrained $1024$ $(4\cdot2^8)$ to $28$ $(3\cdot2^{8-5}+4)$. \Cref{fig:DOFs} compares the \DOFs when using 4D-GSS compared to the unconstrained case of 4D-\abb{GS} for increasing constellation sizes. \Cref{tab:params} shows the \DOFs and the number of 4D shells for a number of different constellation types at a fixed \abb{SE} of $m=8$. This table also shows the amount of discrete 4D energy levels. One of the properties of GSS constellations is the ability to directly and efficiently control the energy of symbols in 4D (due to indexing a power of two shells), which is much more difficult in conventional QAM and GS constellations. 
\section{System Setup and Optimization}\label{sec:sys_setup_optim}

\begin{figure*}[t!]
	\centering
	\begin{adjustbox}{width=1.0\textwidth}
        \includegraphics[]{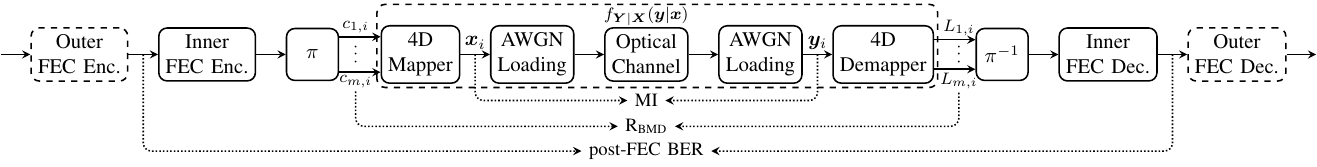}
	\end{adjustbox}
	\caption{Considered 400ZR compatible system with AWGN loading at the transmitter and receiver. The dashed blocks indicate the optional outer SCC FEC and $\pi$ and $\pi^{-1}$ denote the interleaver and inverse interleaver respectively.}
	\label{fig:system_setup_extended}
\end{figure*}

\subsection{Link specification}
For designing constellations with high nonlinear tolerance in mind, a suitable transmission scenario needs to be chosen which is expected to have high \abb{NLIN}. For this reason an unamplified 400ZR link is chosen \cite{OIF-400ZR}, for which the transmitted constellation is PM-16QAM, which matches the desired cardinality of the considered 4D-GSS constellation.

The considered system transmits a dual-polarized, single channel waveform of $2^{20}$ symbols with a fixed random seed over a single span of \abb{SSMF}. This setup uses the Manakov equation as the fiber model \cite[Sec. IV]{Wai1996} and is simulated via the split-step Fourier method (SSFM) with 1000 steps per span using uniform step size and fiber parameters $\alpha = 0.2$~dB/km, $\beta_2=-21.68$~ps$^2$/km and $\gamma = 1.20$~(W$\cdot$km$^{-1}$). Increasing the number of steps per span above 1000 has been verified to provide identical results for the highest launch powers and distances considered in this paper. The symbol rate is matched to the 400ZR specification at 59.84 Gbaud. At the transmitter, the generated symbol sequences are upsampled to 4 samples per symbol and pulse shaped using a root-raised-cosine filter with a roll-off of 1\%. At the receiver, the waveforms are ideally compensated for chromatic dispersion, matched filtered, downsampled to 1 sample per symbol and corrected for phase rotation by utilizing cross-correlation between the input and output symbol sequences. Demapping is done in 4D using \cref{eq:LLR} to calculate the \abbpl{LLR}.

Since an unamplified link is used, noise sources due to transmitter and receiver impairments are emulated instead of simulating optical \abb{EDFA} amplification. For this, worst-case design parameters from the 400ZR specification will be used as a guideline.

\subsection{Transceiver impairments} 
At the transmitter side, the 400ZR specification requires the in-band \abb{OSNR} to have a minimum value of 34 dB/0.1nm. At the receiver side, the concatenated \abb{FEC} scheme in the 400ZR standard is specified to operate error-free (post-FEC BER = $10^{-15}$) when a pre-FEC BER of $1.25\cdot10^{-2}$ or lower is achieved. The receiver sensitivity requires least $-20$ dBm of power to be present at the input of the receiver. These previous two conditions combined with the minimum of 34 dB \abb{OSNR} at the transmitter guarantee error-free operation.

In simulations, the transmitter is emulated by adding \abb{AWGN} to the transmitted waveform such that the \abb{OSNR} value is equal to the in-band \abb{OSNR} limit. For the receiver it is possible to calculate the necessary \abb{AWGN} addition using \cite[Eq. (18)]{Lu1999}
\begin{equation}
    \small
     P_e \cong \frac{4}{m}\left(1-\frac{1}{\sqrt{M}}\right)\sum_{i=1}^{\sqrt{M}/2}Q\left((2i-1)\sqrt{3\frac{E_b}{N_0}\frac{m}{(M-1)}}\right)
    \label{eq:BER_estimation}
\end{equation}
where $P_e$ is the targeted \abb{BER}, $Q(\cdot)$ is the Q-function and $E_b/N_0$ is the accompanying SNR per bit under a Gray-coded $M$-QAM assumption in an \abb{AWGN} channel. For 16QAM and a \abb{BER} of $1.25\cdot10^{-2}$, \cref{eq:BER_estimation} provides an $E_b/N_0$ of 7.53 dB, which  translates to a \abb{SNR} of 13.5 dB. Assuming that an input power of $-20$ dBm is present at the receiver in a back-to-back scenario, the amount of noise power added in the receiver is equal to $-33.5$ dBm. This is added as \abb{AWGN} after simulating the optical fiber.

\subsection{Forward Error Correction}\label{subsec:FEC}
400ZR uses a concatenated \abb{FEC} scheme as defined in \cite[Sec. 10]{OIF-400ZR} consisting of an outer \abb{SCC} with \abb{HD} decoding of rate $0.937$, and an inner Hamming code with \abb{SD} decoding of rate $0.930$, resulting in a total overhead of $14.8\%$. The \abb{SCC} in 400ZR is defined to be taken from \cite[Annex A]{G_709_2}, which describes a $(255,239)$ \abb{SCC} with blocks of size $512\times510$ and a $(1022,990)$ \abb{BCH} code as the component code. The \abb{FEC} code is a double-extended $(128,119)$ Hamming code using a parity-check matrix as described in \cite[Sec. 10.5]{OIF-400ZR}.

Instead of implementing the full concatenated \abb{FEC} as described above, in this paper we only implemented the Hamming code and a \abb{SD} decoder based on a Chase-I decoder \cite{Chase1972}. A post-FEC \abb{BER} after the Hamming decoder of $4.5\cdot10^{-3}$ is targeted, which is the required pre-FEC BER for the $(255,239)$ \abb{SCC} to achieve a BER of $10^{-15}$ at the output. The scrambling and interleaving steps
are approximated by using a bit-wise fixed random permutation after FEC encoding and the inverse operation before FEC decoding. The full system diagram is shown in \cref{fig:system_setup_extended}.

\subsection{Optimization}\label{subsec:optimization}
In \cref{eq:argmax_1}, the optimization problem for determining $\Xbb^*$ was defined. If the constraints from \cref{sec:4D-GSS} are applied, the optimization only needs to be performed for the 28 resulting \DOFs. To enforce the shell constraints, each point out of the 8 points in $\Xcal_\GSS$ is now represented in spherical coordinates $(r_i,\theta_j,\phi_j,\omega_j)$, where $i=1,2,3,4$ and $j=1,2,\ldots,8$. The optimization problem can now be defined as
\begin{equation}\label{eq:optim_simplified}
        \{\bm{r}^*,\bm{\theta}^*,\bm{\phi}^*,\bm{\omega}^*\} = \underset{\mathclap{\substack{\bm{r}~:~0\leq r_i\leq1 \\ \bm{\theta},\bm{\phi},\bm{\omega}~:~0\leq{\theta_j},{\phi_j},{\omega_j}\leq\pi/2}}}{\text{argmax}}~\RBMD(\bm{r},\bm{\theta},\bm{\phi},\bm{\omega})
\end{equation}
where the parameters $(\bm{r},\bm{\theta},\bm{\phi},\bm{\omega})$ are constrained such that the corresponding points are in $\Xcal_\GSS$.

Since there are no existing constellations which strictly adhere to the chosen constraints, selecting an initialization for the optimization procedure is not straightforward. It was determined on a trial-and-error basis that there were no clear differences in resulting performance after optimization between randomly initialized constellations and constellations which were initialized with a distinct structure. For that reason it was chosen to initialize all parameters at the halfway point between the upper and lower bounds, which are shown in \cref{eq:optim_simplified}.

Optimization over the system in \cref{fig:system_setup_extended} is performed using a patternsearch optimizer \cite{Audet2002}, which is a derivative-free multidimensional optimization algorithm.
Patternsearch automatically finds the optimal way to spread out the constellation. During optimization, the number of transmitted symbols is lowered to $2^{18}$ using a different random seed compared to the results generation. The number of steps per span for the \abb{SSFM} is kept at 1000. To enhance stability during the optimization procedure, fixed random seeds are used for the sequence generation and AWGN noise additions.
\section{Results}\label{sec:results}

\begin{figure*}[!t]
    \centering
    \includegraphics[]{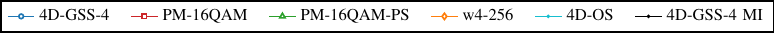}
	\begin{subfigure}[t]{0.55\textwidth}
		\centering
        \includegraphics[]{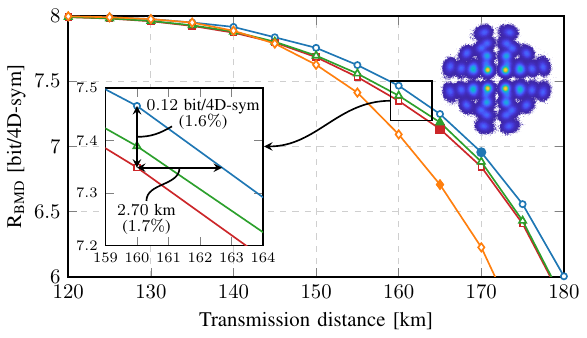}
		\vspace*{-3.5ex}
		\caption{}
		\label{fig:results_RBMD_dist}
	\end{subfigure}%
	\begin{subfigure}[t]{0.45\textwidth}
		\centering
        \includegraphics[]{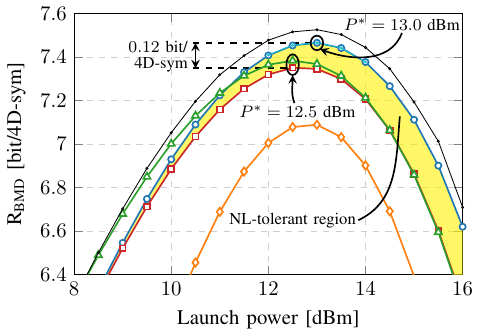}
		\vspace*{-3.5ex}
		\caption{}
		\label{fig:results_RBMD_pow}
	\end{subfigure}%
\caption{$\RBMD$ vs transmission distance at optimal launch power (left) and launch power for a distance of $160$~km (right). Solid markers show the transition to pushing above optimal launch power. Constellation inset shows received symbols at optimal launch power for a distance of $120$~km.}
\label{fig:results_RBMD}
\end{figure*}

\subsection{Baselines}
Conventional PM-16QAM, as used in the 400ZR standard, is considered as a baseline. Next to that, \abb{PS} is applied on top of PM-16QAM by shaping each real dimension with an ideal amplitude shaper, where amplitudes are randomly drawn from a predefined distribution, which is identical over all dimensions. The distribution is optimized for each pair of launch power and transmission distance. This optimization is performed using the same patternsearch optimizer as \cref{subsec:optimization}. Since PM-16QAM only has two amplitude values per dimension, the optimization effectively only has a single \DOF. The proposed 4D-GSS-4 constellations are also optimized and evaluated for each pair of launch power and distance. 
Lastly, a 4D sphere packed constellation is also considered, specifically the 256 point Welti constellation (w4-256) \cite{Welti1974}. Sphere packed constellations are optimal for the \abb{AWGN} channel for a given \abb{SE} and will provide insight into the maximum expected performance in the linear region later in this paper. For $\RBMD$ evaluation, w4-256 uses the optimized binary labeling from \cite{Binary_Labeling}.

It is well known that symbol-level interleaving, as described in one of the \abb{DSP} steps in the 400ZR standard \cite[Sec. 11.1]{OIF-400ZR}, has a negative impact on the performance of \abb{PS} constellations implemented with a finite blocklength amplitude shaper \cite{Fehenberger2020_2}. This is due to symbol-level interleaving breaking time-domain structures of probabilistically shaped symbol sequences and can be mitigated by employing extra pre- and post-interleaving steps in the \abb{DSP} chain \cite{Peng2020}. In this paper, the coded bits of all considered constellations are independent. PM-16QAM-PS uses an ideal amplitude shaper, which does not implement an actual shaping algorithm, resulting in independent bits. 4D-GSS and the two other baselines all use uniform signaling and are thus also not affected by such dependencies. Since all considered constellations are not affected by symbol-level interleaving, only a bit-wise permutation as discussed in \cref{subsec:FEC} was implemented.

\subsection{$\RBMD$ optimized constellations}\label{sec:results_RBMD}
\Cref{fig:results_RBMD_dist} shows $\RBMD$ results for a distance sweep between 120 and 180 km. In the inset of \cref{fig:results_RBMD_dist}, 4D-GSS-4 achieves $1.6\%$ gain in $\RBMD$ and $1.7\%$ gain in distance compared to PM-16QAM around 160km. These gains represent the increase in data rate and distance as a result of the shaping gain. Since in this specific scenario, the 400ZR link is loss-limited, results are shown at optimal launch power until the distance is too large to satisfy the received power requirement of at least $-20$~dBm. The point after which this occurs is indicated by solid markers. Beyond these markers, the constellations are pushed to launch powers above the optimal value to satisfy the received power requirement. In \cref{fig:results_RBMD_dist}, it is shown that the distance for which 4D-GSS-4 can operate at optimal launch power is approximately $5$~km larger than the other considered constellations. This increase in maximum optimal launch power is also reflected in \cref{fig:results_RBMD_pow} where 4D-GSS-4 has $0.5$ dB higher optimal launch power ($P^*$) compared to the baselines and hence, the highest nonlinear tolerance among the considered schemes. The region where NL tolerance is observed is indicated in yellow. In the linear domain, 4D-GSS-4 has similar performance to PM-16QAM and loses in performance compared to PM-16QAM-PS. Even though an optimized binary labeling is used for the w4-256 constellation, since it is not designed to maximize $\RBMD$, it performs poorly.

To evaluate the performance losses induced by the GSS constraints, optimized 4D constellations which have only orthant symmetry as the constraint (denoted with 4D-OS) are evaluated around the optimal launch power. It is shown in \cref{fig:results_RBMD_pow} that removing the shell constraints and X-Y constraint from 4D-GSS-4 has a negligible impact on performance. Furthermore, it has been shown in \cite[Fig. 6]{Chen2021} that lifting the orthant symmetry constraint has a marginal impact on performance. All this indicates that the proposed symmetry constraints as used in 4D-GSS-4 have very little impact on total performance, but do reduce the optimization complexity significantly. \Cref{fig:results_RBMD_pow} also includes \abb{MI} results for 4D-GSS-4 (denoted by 4D-GSS-4 MI). This indicates the theoretical upper limit for the $\RBMD$ where it is clear that quite a large gap still exists between the $\RBMD$ and the \abb{MI} for 4D-GSS-4.

\begin{figure}
    \centering
    \includegraphics[]{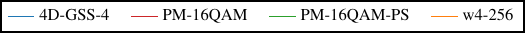}
	\begin{subfigure}[t]{0.5\linewidth}
		\centering
        \includegraphics[]{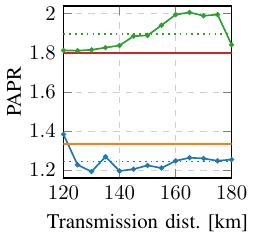}
		\caption{}
		\label{fig:4D_PAPR_GMI}
	\end{subfigure}%
	\begin{subfigure}[t]{0.5\linewidth}
		\centering
        \includegraphics[]{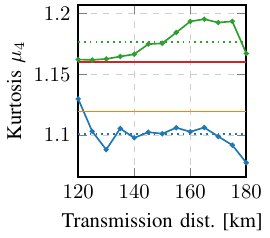}
		\caption{}
		\label{fig:4D_kurtosis_GMI}
	\end{subfigure}%
	\caption{PAPR (linear) (left) and 4D kurtosis $\mu_4$ (right) for $\RBMD$ optimized constellations at different transmission distances and optimal launch power. Dotted lines denote average values (arithmetic mean) over the shown distances.}
    \label{fig:PAPR_kurtosis_GMI}
\end{figure}

Possible explanations of the increased nonlinear tolerance of 4D-GSS-4 can be observed from \cref{fig:PAPR_kurtosis_GMI}, which shows the \abb{PAPR} (in linear units) and the fourth order standardized moment (i.e. the kurtosis\footnote{The kurtosis of a distribution depends on the dimensionality \cite[Sec. III]{Rosiers1999}, where maximum kurtosis is achieved for a Gaussian distribution. Maximum kurtosis values for 2D and 4D constellations are $2$ and $1.5$ respectively. In this paper, only 4D kurtosis values are shown. It is also common practice to compare the kurtosis of an $N$-D distribution to that of an $N$-D univariate normal distribution. This is typically done by defining `excess kurtosis', which is the kurtosis of the distribution minus the kurtosis of the Gaussian distribution, then comparing it to zero.}) of the considered constellations. \abb{PAPR} is a rough indicator for evaluating nonlinear tolerance \cite[Sec. II-D]{Chen2021} and also influences the amount of distortion resulting from limited equipment dynamic range and linearity, especially in orthogonal frequency division multiplexing transmission \cite[Sec. V-A]{Armstrong2009}. Kurtosis is considered to be more directly related to nonlinearities, where channel input distributions with high kurtosis lead to higher \abb{NLIN} power \cite[Sec. III-B]{Fehenberger2016}. \Cref{fig:4D_PAPR_GMI} shows that PM-16QAM has a \abb{PAPR} of 1.8, while 4D-GSS-4 constellations have a \abb{PAPR} of 1.25 on average over the considered distances, which is a reduction of $31\%$ compared to PM-16QAM. Similarly, \Cref{fig:4D_kurtosis_GMI} shows a decrease in the kurtosis from 1.16 for PM-16QAM to 1.10 for 4D-GSS-4, which is a reduction of $5\%$.

\begin{figure}[]
    \centering
    \includegraphics[]{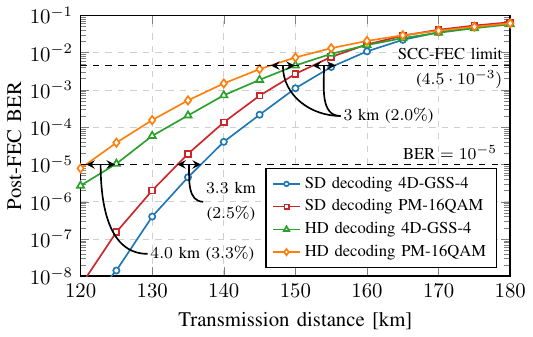}
    \caption{Post-FEC BER vs transmission distance at optimal launch power for a $(128,119)$ Hamming code with SD and HD decoding. SCC-FEC limit denotes the maximum BER at the output of the Hamming codes for error free operation of the outer staircase code.}
    \label{fig:BER_Hamming}
\end{figure} 

\Cref{fig:BER_Hamming} shows the post-FEC BER of 4D-GSS-4 compared to PM-16QAM when the inner Hamming code is \abb{SD}-decoded as described in \cref{subsec:FEC}, together with \abb{HD} decoding of the same code. The outer SCC has a FEC limit of $4.5\cdot10^{-3}$, which is used as the minimum required BER after the Hamming code for error-free operation. A gain of $2\%$ in transmission distance between 4D-GSS-4 and PM-16QAM is achieved at the SCC-FEC limit, which is very close to the observed $1.6\%$ gain in $\RBMD$. The HD-decoded Hamming code shows similar gains between the two constellation types but cannot satisfy the SCC-FEC limit over similar distances. When only the Hamming codes are considered without the outer SCC, gains increase to in-between $2.5\%$ and $3.3\%$ depending on the specific distance and code, as indicated by the $10^{-5}$ BER line.

\subsection{MI optimized constellations}\label{subsec:MI_opt_const}

\begin{figure*}[!ht]
    \centering
        \includegraphics[]{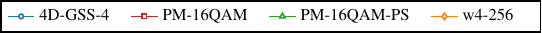}
	\begin{subfigure}[t]{0.55\textwidth}
		\centering
        \includegraphics[]{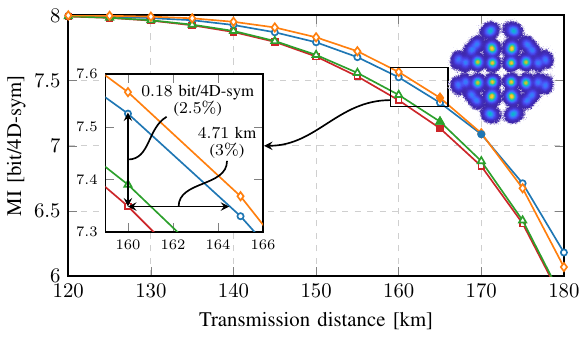}
		\vspace*{-3.5ex}
		\caption{}
		\label{fig:results_MI_dist}
	\end{subfigure}%
	\begin{subfigure}[t]{0.45\textwidth}
		\centering
        \includegraphics[]{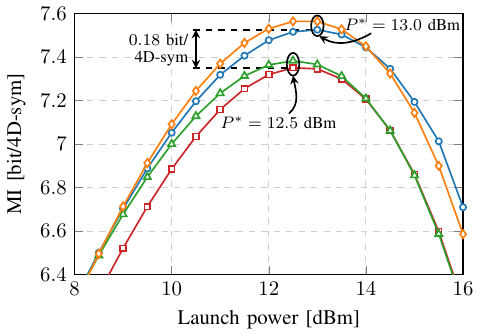}
		\vspace*{-3.5ex}
		\caption{}
		\label{fig:results_MI_pow}
	\end{subfigure}%
\caption{MI vs transmission distance at optimal launch power (left) and launch power for a distance of $160$~km (right). Solid markers show the transition to pushing above optimal launch power. Constellation inset shows received symbols at optimal launch power for a distance of $120$~km.}
\label{fig:results_MI}
\end{figure*}

\begin{figure}
    \centering
    \includegraphics[]{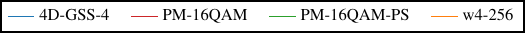}
	\begin{subfigure}[t]{0.5\linewidth}
		\centering
        \includegraphics[]{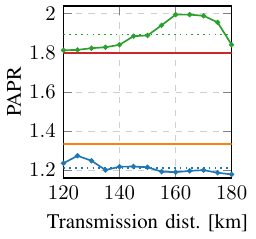}
		\caption{}
		\label{fig:4D_PAPR_MI}
	\end{subfigure}%
	\begin{subfigure}[t]{0.5\linewidth}
		\centering
        \includegraphics[]{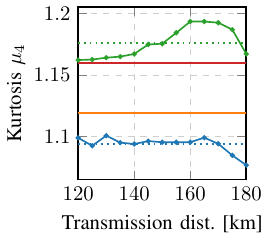}
		\caption{}
		\label{fig:4D_kurtosis_MI}
	\end{subfigure}%
	\caption{PAPR (linear) (left) and 4D kurtosis $\mu_4$ (right) for MI optimized constellations at different transmission distances and optimal launch power. Dotted lines denote average values (arithmetic mean) over the shown distances.}
    \label{fig:PAPR_kurtosis_MI}
\end{figure}

It was observed in \cref{fig:results_RBMD_pow} that the $\RBMD$ for the optimized 4D-GSS-4 constellations was consistently lower than the \abb{MI} by about $0.06$ bits/4D-sym. This could indicate an issue with the binary labeling since PM-16QAM(-PS) did not show such a gap. To find the potential upper performance limit of 4D-GSS-4, the optimization procedure was repeated using the \abb{MI} as the performance metric using \cref{eq:MI} for evaluating the \abb{MI}.
This results in the following optimization problem
\begin{equation}
    \Xbb^* = \underset{\Xbb\in\Xcal_{\GSS}}{\argmax}~\text{MI}(\Xbb).
    \label{eq:argmax_1}
\end{equation}

\Cref{fig:results_MI_dist} shows MI results for a distance sweep between 120 and 180 km. The inset of \Cref{fig:results_MI_dist} shows gains of $3\%$ in distance and $2.5\%$ in MI for 4D-GSS-4 vs. PM-16QAM and is slightly outperformed by w4-256. Again, \cref{fig:results_MI_pow} shows that 4D-GSS-4 has the largest optimal launch power. Moreover, due to the rapidly-vanishing MI of w4-256, 4D-GSS-4 outperforms w4-256 at very high powers ($P>14$). However, for lower powers ($P<14$), w4-256 achieves larger MI than 4D-GSS-4. The observed gap in performance for 4D-GSS-4 when comparing MI to $\RBMD$ indicates a possible issue where the proposed \abb{GSS} framework does not provide a structure suitable for a good binary labeling.

Results in \cref{fig:PAPR_kurtosis_MI} show similar trends to \cref{fig:PAPR_kurtosis_GMI}, with the main difference being that the MI optimized 4D-GSS-4 constellations have even lower average PAPR and kurtosis. The difference in PAPR against PM-16QAM increases from $31\%$ to $33\%$, whilst the difference in kurtosis increases from $5\%$ to $6\%$. Against w4-256, 4D-GSS-4 shows a reduction in PAPR of $10\%$ and a reduction in kurtosis of $2\%$, which could contribute to 4D-GSS-4 gaining performance in terms of MI over w4-256 for launch powers larger than $14$ dBm.

\subsection{Bitwise MI}
\begin{figure}
    \centering
    \includegraphics[]{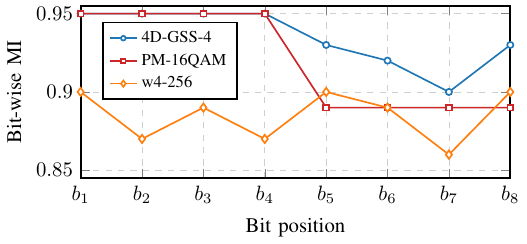}
    \caption{Bit-wise MI results at 160km and optimal launch power.}
    \label{fig:bitwise_MI}
\end{figure}

To investigate possible causes of the binary labeling penalty for 4D-GSS-4, we look at the bit-wise MI $I(C_k;\bmY)$. \Cref{fig:bitwise_MI} compares the bit-wise MI of 4D-GSS-4, PM-16QAM and w4-256 at the optimal launch powers. The individual bits are denoted by $b_i$ for $i=1,\ldots,m$. For PM-16QAM, the bits are reordered such that the bits which determine the signs are the first four bits (same as 4D-GSS-4). This does not effect the performance since PM-16QAM is a Cartesian product of four independent \abb{PAM}-4 constellations. As a result, this also implies symmetry across all four dimensions and thus, PM-16QAM also has orthant symmetry.

In the binary labels, the bits $b_1$ through $b_4$ determine the orthant for both PM-16QAM and 4D-GSS-4. The bits $b_5$ through $b_8$ determine the amplitude of each of the PAM-4 signals for PM-16QAM. For 4D-GSS-4, the bits $b_5$ and $b_6$ select the shell, bit $b_7$ selects between 2 points on a shell, and bit $b_8$ selects between X-Y symmetric points.

In terms of bit-wise MI, $b_1$ through $b_4$ perform identically for both PM-16QAM and 4D-GSS-4, which is expected due to both constellations employing orthant symmetry. Bits $b_5$ trough $b_8$ have larger bit-wise MI for 4D-GSS-4 compared to PM-16QAM, where bit $b_7$ has the lowest value. This suggests that the proposed structure combined with the chosen constellation cardinality does not allow for a good labeling of $b_7$.
Lastly, as expected, w4-256 has much worse performance in general compared to the other two constellations since the structure of this constellation is not optimized for allowing a good binary labeling at all.
\section{Conclusion}\label{sec:conclusions}
A novel framework is proposed for generating families of well-structured 4D geometrically-shaped constellations which are more nonlinearity-tolerant than conventional PM-16QAM. Numerical simulations show that the newly proposed 4D-GSS-4 constellations outperform both PM-16QAM and PS-PM-16QAM in a 400ZR-compatible transmission setup when optimized for $\RBMD$. It was shown that the imposed constraints lead to negligible performance degradation while considerably reducing the optimization space and resulting in well-structured constellations. It was also found that the chosen constraints combined with the chosen \abb{SE} do not allow for very good binary labeling, which is indicated by optimizing 4D-GSS-4 for \abb{MI}, which resulted in a reach increase of $3\%$.

Investigating better combinations of constellation cardinality (${\geq10}$ bits/4D-sym) and GSS (e.g., modifying shell constraints) is left for further investigation. Another area of possible research is to increase the dimensionality across channel time slots or number of wavelength channels (e.g., 8D-GSS).


\end{document}